\documentclass{revtex4}

\usepackage{graphicx}
\usepackage{amsmath}
\usepackage{amssymb}
\usepackage{color}

\begin{document}

\title{Noise-induced drift in stochastic differential equations with arbitrary friction and diffusion in the Smoluchowski-Kramers limit \thanks{S.H. was supported by the VIGRE grant through the University of Arizona Applied Mathematics Program.
J.W. was partially supported by the NSF grant DMS 1009508.}}


\author{Scott Hottovy}
\author{Jan Wehr}
\affiliation{Department of Mathematics and Program in Applied Mathematics, University of Arizona, Tucson, Arizona 85721 USA}
\author{ Giovanni Volpe}
\affiliation{Max-Planck-Institut f\"{u}r Intelligente Systeme, Heisenbergstra$\beta$e 3, 70569 Stuttgart, Germany\\
2. Physikalisches Institut, Universit\"at Stuttgart, Pfaffenwaldring 57, 70569 Stuttgart, Germany\\
Department of Physics, Bilkent University, Cankaya, Ankara 06800, Turkey\\}




\begin{abstract}
We consider the dynamics of systems with arbitrary friction and diffusion. These include, as a special case, systems for which friction and diffusion are connected by Einstein fluctuation-dissipation relation, e.g. Brownian motion. We study the limit where friction effects dominate the inertia, i.e. where the mass goes to zero (Smoluchowski-Kramers limit). {Using the It\^o stochastic integral convention,} we show that the limiting effective Langevin equations has different drift fields depending on the relation between friction and diffusion.  {Alternatively, our results can be cast as different interpretations of stochastic integration in the limiting equation}, which can be parametrized by  $\alpha \in \mathbb{R}$. 
Interestingly, in addition to the classical It\^o ($\alpha=0$), Stratonovich ($\alpha=0.5$) and anti-It\^o ($\alpha=1$) integrals, we show that position-dependent $\alpha = \alpha(x)$, and even stochastic integrals with $\alpha \notin [0,1]$ arise.  Our findings are supported by numerical simulations.
\end{abstract}

\maketitle

\section{Introduction}\label{sec: intro}

Most physical, chemical, biological and economic phenomena present an intrinsic degree of randomness. These are typically modelled by stochastic differential equations (SDEs) \cite{oksendal}. SDEs were introduced at the beginning of the XX century to describe Brownian motion by adding a random driving function to an ordinary differential equation (ODE); since then, SDEs have come into widespread use in, e.g., physics, biology, and economics. However, SDEs involving multiplicative noise terms can be integrated according to various definitions leading to different solutions \cite{karatzas}, e.g. the It\^o integral and the Stratonovitch integral. From the modeling point of view it is, therefore, key to know what definition to use in any given situation \cite{vankampen}. {From the mathematical point of view the simplest approach is to write all equations according to, e.g., the It\^o definition of the stochastic integral; the different interpretations mentioned above then reappear as additional drift terms, which are often referred to in the literature as ``spurious drifts."}

In order to understand the origin of the difficulty, we consider a simple case. The SDE $dx_t = \sigma(x_t) dW_t$, where $W_t$ is a Wiener process, can be solved by integration, i.e. $\displaystyle  x_t = x_0 + \int_0^t \sigma(x_s)dW_s$, 
{where we define $\displaystyle \int_0^t \sigma(x_s) \circ^\alpha dW_s = \lim_{N \rightarrow \infty} \sum_0^N \sigma(x_{t_n^\alpha}) \Delta W_{t_n} $ with $\Delta W_{t_n} = W_{t_{n+1}}-W_{t_n}$, $t_n = \frac{nt}{N}$ and $t_n^\alpha = \frac{n+\alpha}{N}t$ and, typically, $\alpha \in [0,1]$}. Since $W_t$ is a function of unbounded variation, differently from ordinary Riemann-Stieltjes integrals, the limit of these partial summations generally leads to different values of the integral depending on the choice of $\alpha$. In particular, $\alpha = 0$ leads to the It\^o integral, $\alpha = 0.5$ leads to the Stratonovitch integral, and $\alpha = 1$ leads to the anti-It\^o (or isothermal) integral. This is the reason why it is necessary to give both a SDE and the respective $\alpha$ with which to solve it in order to have a fully determined model \cite{vankampen}.

Various preferences regarding the appropriate choice of $\alpha$ have emerged in the numerous fields where SDEs have been applied�\cite{oksendal}. For example, the martingale property, i.e. the specific feature of the It\^o integral of ``not looking into the future," meaning that, when the integral is approximated by a summation, the leftmost point of each interval is used, is the main reason of its popularity in economics \cite{oksendal} and biology \cite{Turelli1977,Ao2008}. In general, the Stratonovitch integral emerges naturally when the Wiener process is replaced by a sequence of approximating deterministic processes and has the advantage of leading to ordinary chain rule formulas under a change of variable \cite{gardiner}. However, the fact that Stratonovitch integrals are not martingales gives the It\^o integral an important computational advantage \cite{kloeden,sussmann1978}. Finally, the anti-It\^o integral has been shown to be the most appropriate to describe physical phenomena that are in e
 quilibrium with a heat-bath for which Einstein fluctuation-dissipation relation holds \cite{ermark1978,lancon2001,lau2007,volpe2010,brettschneider2011}. {In particular, equations that satisfy the fluctuation-dissipation relation occur in molecular dynamics. Here, the limiting equation is constrained to be the anti-It\^o type to correct the invariant distribution to model the Gibbs distribution.}

\begin{figure}
\centering
\includegraphics[width=12cm]{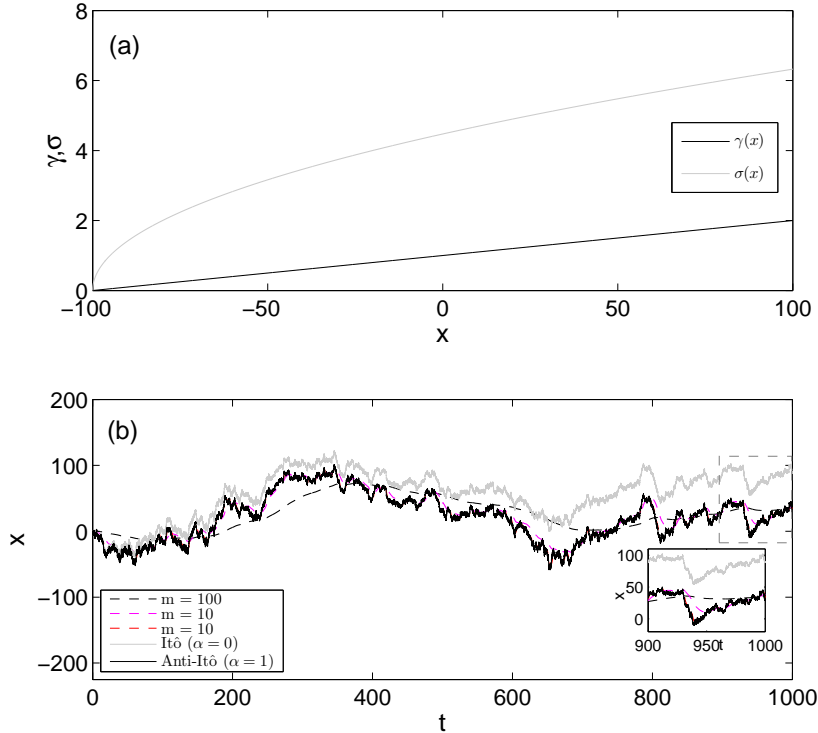}
\caption{(a) For a Brownian particle $\gamma(x)$ (dark line) and $\sigma(x)$ (grey line) are related by the Einstein fluctuation-dissipation relation [Eq. (\ref{eq:einstein})]; in this case, $\gamma(x)=(1+x/100)$ (b) The solution of the Newton equations [Eq. (\ref{eq:LEgeneral})] for $m \rightarrow 0$ (dashed lines) converge to the solution of the SK approximation [Eq. (\ref{eq:SDE})] for $\alpha = 1$, i.e. anti-It\^o integral, (black solid line); also the solution for $\alpha = 0$, i.e. It\^o integral, (grey solid line) is given for comparison. All solutions are obtained for the same Wiener process. The inset is a blow-up of the final part of the trajectories (dashed square).}
\label{fig:alpha1}
\end{figure}

A classical example of a phenomenon in equilibrium with a heat bath is the motion of a mesoscopic particle of mass $m$ immersed in a fluid, i.e. Brownian motion. If we assume that the particle moves in one dimension under the action of a continuous force $F(x)$, its position being $x_t^m \in \mathbb{R}$ at all times $t \geq 0$ in a finite interval, the corresponding Newton equation is:
\begin{equation}\label{eq:LEgeneral}
\left\{\begin{array}{rcl}
dx_t^m & = & v_t^m dt\\
m dv_t^m & = & F(x_t^m) -\gamma(x_t^m)v_t^m dt + \sigma(x_t^m)dW_t
\end{array}\right.
\end{equation} 
with the initial conditions $v_0^m = \nu^m$ and $x_0^m = \chi^m$.
The friction coefficient $\gamma(x)>0$ and the intensity (standard deviation) of the noise $\sigma(x)>0$ are, in general, position-dependent; we also assume that $F$, $\gamma$, and $ \sigma$ are differentiable functions of $x$ smooth enough so that the process $(x_t,v_t)$ exists for all $t$ on a finite interval. It is well known that, since the derivative of $x_t$, i.e. $\dot{x}_t^m = v_t^m$, exists, the stochastic integral in Eq. (\ref{eq:LEgeneral}) is equivalent under all interpretations \cite{oksendal,karatzas,gardiner}. 

{The limit of Eq. (\ref{eq:LEgeneral}) as $m \to 0$ has to be studied with care, requiring a nontrivial computation.
In general, similar limits involve additional drift terms, i.e. ``spurious drifts." A statement of our results in terms of different definitions of stochastic integral is also possible and, in some cases, straightforward.} By naively setting $m=0$ in Eq. (\ref{eq:LEgeneral}), we obtain a SDE for $x_t^0=x_t$:
\begin{equation}\label{eq:SDE}
dx_t = \underbrace{\frac{F(x_t)}{\gamma(x_t)}\;dt}_{\mbox{deterministic}} + \underbrace{\frac{\sigma(x_t)}{\gamma(x_t)}\;dW_t}_{\mbox{stochastic}},
\end{equation}
with initial condition $x_0 = \chi^0$.
Eq. (\ref{eq:SDE}) is called the Smoluchowski-Kramers (SK) approximation to Eq. (\ref{eq:LEgeneral}).
Differently from the solution of Eq. (\ref{eq:LEgeneral}), the solution of Eq. (\ref{eq:SDE}) depends on the interpretation of the stochastic term, i.e. on the choice of $\alpha$. 
{In the following we identify the value of $\alpha$ that introduces the correct additional drift; we emphasize that this is just another way of expressing additional drift terms within the It\^o formalism.}

We can gain some insight into this zero-mass limit procedure considering numerical solutions of Eq. (\ref{eq:LEgeneral}) for various decreasing values of $m$, but for the same Wiener process [Fig. \ref{fig:alpha1}].
For a Brownian particle the Einstein fluctuation-dissipation relation holds:
\begin{equation}\label{eq:einstein}
\gamma(x) \propto \sigma(x)^2.
\end{equation}
In Fig. \ref{fig:alpha1}(a) $\gamma(x)$ (dark line) and $\sigma(x)$ (grey line) are presented. The dashed lines in Fig. \ref{fig:alpha1}(b) represent some solutions of Eq. (\ref{eq:LEgeneral}) for decreasing $m$: they become rougher and rougher as the $m$ decreases. They converge towards the anti-It\^o ($\alpha = 1$) solution of Eq. (\ref{eq:SDE}) (black solid line); this is in agreement with the recent experimental demonstration \cite{volpe2010,brettschneider2011} and mathematical proof \cite{wehr2011} that for a Brownian particle the most natural interpretation is the anti-It\^o integral.
We remark that the It\^o ($\alpha = 0$) solution of Eq. (\ref{eq:SDE}) (grey solid line) presents clear deviations from the correct one, as can be clearly seen in the inset of Fig. \ref{fig:alpha1}(b).

In this article we study  the zero-mass limiting behavior of a larger class of equations that have the form of Eq. (\ref{eq:LEgeneral}), but for which $\gamma(x)$ and $\sigma(x)$ are allowed to vary independently from each other. This can be the case, e.g., in the description of the evolution of complex systems \cite{Ao2007}. We study this general class of equations from the point of view of the convergence of the infinitesimal operators of the corresponding diffusion processes, i.e. backward Kolmogorov equations, thus following the well-known methods from homogenization theory \cite{papanicolaou1975,pavliotis,schuss}. The main result of this paper identifies for given $\gamma(x)$ and $\sigma(x)$ the drift term and hence the corresponding $\alpha$ [Eq. (\ref{eq:alpha})]: we find that in general it can be a function of $x$. {We remark that the spurious drifts are defined assuming the It\^o stochastic calculus convention, while the values of $\alpha$ are defined with reference to SDE (\ref{eq:SDE}); we introduce the notation $\circ^{\alpha(x)}dW_t$ to indicate the presence of such extra drift.}
Interestingly, we find that when a generalized fluctuation-dissipation relation holds, i.e.
\begin{equation}\label{eq:generalized}
\gamma(x) \propto \sigma(x)^{\lambda},
\end{equation}
$\alpha$ is only a function of the exponent $\lambda$ and independent from $x$ [Eq. (\ref{eq:alpha_lambda})]. In particular, for $\lambda = 0$ we retrieve the It\^o interpretation and for $\lambda = 2$ the anti-It\^o interpretation, while the Stratonovich interpretation is only retrieved asymtotically for $\lambda \rightarrow \infty$. Interestingly, also values of $\alpha \notin [0,1]$ occur for $\lambda \in (0,2)$. {Although in this article we always consider the variable $x$ to be one dimensional, the general multi-dimensional case can be studied using similar methods \cite{pavliotis}.}

In section \ref{sec:model}, we give further details on the ansatz we will use. In section \ref{sec:modified}, we introduce a relation [Eq. (\ref{eq:SKsdeItox_general}) and Eq. (\ref{eq:SKsdeItox})] that permits one to express any stochastic integral as an It\^o integral with a modified drift term. In section \ref{sec:SKapprox}, we analyze an asymptotic expansion of the solution of the Kolmogorov equation in order to derive the effective SK-approximation [Eq. (\ref{eq:SDEgenericSK})]. In section \ref{sec:alpha}, we derive a formula for the value of $\alpha(x)$ as a function of $\gamma(x)$ and $\sigma(x)$ [Eq. (\ref{eq:alpha})] and we study various important special cases: the case of constant $\gamma(x)$ [$\mathsection$ \ref{ssec:constant}]; the case of a Brownian particle for which Eq. (\ref{eq:einstein}) holds [$\mathsection$ \ref{ssec:brownian}]; the cases for which Eq. (\ref{eq:generalized}) holds, which are all the cases where $\alpha$ is position-independent [$\mathsection$ \ref{ssec:specialcases}]; and the singular case for which $\gamma(x) \propto \sigma(x)$, which leads to a SDE without ambiguity, but with an additional drift nevertheless [$\mathsection$\ref{ssec:proportional}].

\section{The ansatz}\label{sec:model}

Our ansatz is to study the zero-mass limit of a diffusion process defined by Eq. (\ref{eq:LEgeneral}) with arbitrary $\gamma(x)$ and $\sigma(x)$, analyzing the behavior of its infinitesimal generator, i.e. the Kolmogorov equation. We perform an asymptotic analysis of the Kolmogorov equation, expanding its solution in powers of a small parameter, i.e. $\sqrt{m}$. The result is a drift term in the effective SDE, which can be translated into the correct interpretation of the stochastic integral. In order to clearly present the results in a technicality-free style, the approach is not fully rigorous. {Similar convergence results have been studied \cite{Pardoux2003}}. 

The problem of taking the mass to zero, justifying the limiting equation, has been addressed by various authors at different levels of generality and mathematical rigor, starting with M. Smoluchowski \cite{smoluchowski1916} and H. Kramers \cite{kramers1940}. E. Nelson studied the $F=0$ case in which $\gamma$ and $\sigma$ were constant and proved that the solution of Eq. (\ref{eq:LEgeneral}) converges to the solution of Eq. (\ref{eq:SDE}), which in this case was unambiguous since $\sigma$ was constant \cite{nelson}.  The case with constant $\sigma$ but including an external force was also treated (see reference \cite{schuss} and references therein) with a similar result but by entirely different methods. The case where $\sigma$ depended on the position, but $\gamma$ was constant, was first studied by M. Freidlin in \cite{freidlin2004}, showing that the limiting equation should be interpreted with $\alpha = 0$ and that, in the presence of the colored noise, in some cases the limiting $\alpha
 $ equals ${1 \over 2}$.  Subsequently, R. Kupferman \emph{et al.} showed that all values of $\alpha$ between $0$ and ${1 \over 2}$ could be obtained by taking the correlation of the noise and the mass of the particle to zero in an appropriate way \cite{kupferman2004} .

\section{Stochastic integrals as modified It\^o integrals}\label{sec:modified}

A stochastic integral with a given $\alpha(x)$ can be expressed as an It\^{o} integral, i.e. $\alpha = 0$, with an additional noise-induced drift, {i.e. a ``spurious drift."} To justify this claim, we consider an arbitrary process
\begin{equation*}
dx_t = b(x_t) dt + \sigma(x_t)\circ^{\alpha(x)} dW_t,
\end{equation*}
where the SDE is defined by $\alpha(x)$ {and the solution $x_t$, is real valued and one dimensional}. The integrated equation is
\begin{equation*}
x_t = x_0 + \int_0^t b(x_s)\;ds + \lim_{N\rightarrow\infty}\sum_{n=0}^N \sigma(x_{t_n^\alpha})\Delta W_{t_n},
\end{equation*}
with $t_n^\alpha = \frac{n+\alpha(x_{t_n})}{N}t$. By expanding $\sigma(x_t)$, we see that this corresponds to
\begin{equation*}
x_t = x_0 + \int_0^t b(x_s)\;ds + \int_0^t \alpha(x_s) \sigma(x_s)\frac{d\sigma(x_s)}{dx_s}\;ds + \lim_{N\rightarrow\infty}\sum_{n=0}^N \sigma(x_{t_n})\Delta W_{t_n}, 
\end{equation*}
with $t_n = \frac{n}{N}t$. And this can be interpreted as the It\^o ($\alpha \equiv 0$) equation
\begin{equation}\label{eq:SKsdeItox_general}
dx_t = b(x_t) dt + \alpha(x_t) \sigma(x_t) \frac{d\sigma(x_t)}{dx_t} + \sigma(x_t)dW_t,
\end{equation}
{where we omit the $\circ^0$ for all further It\^o integrals.}
In particular, Eq. (\ref{eq:SDE}) interpreted with any $\alpha(x)$ corresponds to the It\^{o} equation:
\begin{equation}\label{eq:SKsdeItox}
dx_t = \left [\frac{F(x_t)}{\gamma(x_t)} + \alpha(x_t) \frac{\sigma(x_t)}{\gamma(x_t)}\frac{d}{dx_t}\left ( \frac{\sigma(x_t)}{\gamma(x_t)}\right )\right ]\;dt + \frac{\sigma(x_t)}{\gamma(x_t)}\;dW_t,
\end{equation}
with initial condition $x_0 = \chi^0$.

\section{Smoluchowski-Kramers approximation}\label{sec:SKapprox}

In order to simplify further analysis, we substitute $u_t^m = \sqrt{m}v_t^m$ in Eq. (\ref{eq:LEgeneral}) obtaining the following two-dimensional SDE:
\begin{equation}\label{eq:SDEgeneral}
\left\{\begin{array}{rcl}
dx_t^m & = & \frac{1}{\sqrt{m}}u_t^m\;dt \\
du_t^m & = & \left[ \frac{F(x_t^m)}{\sqrt{m}} - \frac{\gamma(x_t^m)}{m}u_t \right] \;dt + \frac{\sigma(x_t^m)}{\sqrt{m}}\;dW_t
\end{array}\right.
\end{equation} 
with initial conditions $x_0^m = \chi^m$ and $u_0^m = \sqrt{m}\nu^m$.

To determine $\alpha$ we use a multiscale analysis of the backward Kolmogorov equation of the SDE (\ref{eq:SDEgeneral}).  Let $g(x',u',t'|x,u,t)$ be the probability density of the distribution of the position and (rescaled) velocity $(x', u')$
of the particle at time $t'$ given their values $(x,u)$ at a time $t<t'$. Then the backward Kolmogorov equation for the SDE (\ref{eq:SDEgeneral})  is
\begin{align}\label{eq:BKgeneralSDE}
\frac{\partial g(x',u',t'|x,u,t)}{\partial t} &= \frac{\sigma(x)^2}{2m}\frac{\partial^2 g(x',u',t'|x,u,t)}{\partial u^2} + \frac{u}{\sqrt{m}}\frac{\partial g(x',u',t'|x,u,t)}{\partial x} \\
    +& \frac{F(x)}{\sqrt{m}}\frac{\partial g(x',u',t'|x,u,t)}{\partial u} - \frac{\gamma(x)u}{m}\frac{\partial g(x',u',t'|x,u,t)}{\partial u}. \nonumber
\end{align}
Since the equation involves derivatives with respect to the $x, u,$ and $t$ variables we write $g(x',u',t'|x,u,t) = g(x,u,t)$, to shorten notation. Eq. (\ref{eq:BKgeneralSDE}) can be rewritten as
\begin{equation}
	\label{eq:Loperators}
	\frac{\partial g}{\partial t} = \left (\frac{1}{m}L_1 +\frac{1}{\sqrt{m}}L_2\right )g.\nonumber
\end{equation}
with 
\begin{align}
    L_1 &= \frac{\sigma(x)^2}{2}\frac{\partial^2}{\partial u^2} - \gamma(x) u\frac{\partial}{\partial u},\nonumber \\
    L_2 &= u\frac{\partial }{\partial x} + F(x) \frac{\partial }{\partial u}.\nonumber
\end{align}
{Notice that the operator $L_1$ is the generator for an Ornstein-Uhlenbeck (OU) process with coefficients dependent on $x$. We denote this process as $\tilde{u}$, and write the stochastic differential equation
\begin{equation}
	d\tilde{u}_t = -\gamma(x)\tilde{u}_t\;dt + \sigma(x)\;dW_t. 
\end{equation}
The invariant density for $\tilde{u}_t$ is 
\begin{equation}
	\label{eq:invdensity}
	g^*(\tilde{u})= C(x) \exp\left (\frac{-\gamma(x)\tilde{u}^2}{\sigma(x)^2}\right ),
\end{equation}
where $C(x)$ is a normalizing constant. 
} 
We postulate that the solution of the Kolmogorov equation has an asymptotic expansion $g = g_0 + \sqrt{m}g_1 + mg_2 + ...$ \cite{papanicolaou1975,pavliotis,schuss}. We match powers of $m$  to obtain the following equations, 
\begin{align}
	\label{eq:m}
    L_1g_0 &= 0, \\
	\label{eq:sqrtm}
    L_1g_1 &= -L_2g_0, \\
	\label{eq:unity}
    \frac{\partial g_0}{\partial t} &= L_1g_2 + L_2g_1, 
\end{align}
where $L_1,L_2$ are differential operators. 
Solving Eq. (\ref{eq:m}) results in
\begin{equation}
    g_0(x,u,t) = C_1(x,t)\int_{-\infty}^u e^{\frac{\gamma(x)\hat{u}^2}{\sigma(x)^2}}\;d\hat{u} + C_2(x,t). \nonumber
\end{equation}
  Since the first term is not integrable in $u$, $C_1$ must be zero and thus $g_0 = g_0(x,t)$, independent of $u$.  By the Fredholm alternative, the solvability condition for Eq. (\ref{eq:sqrtm}) is given as
\begin{equation}
	\int_{-\infty}^\infty g^*L_2g_0\;du = 0, \nonumber
\end{equation}
for all $g^*$ such that $L_1^*g^*=0$ \cite{courant,lax} .  Here $L_1^*$ is the adjoint of $L_1$: 
\begin{equation}
	L_1^* = \frac{\sigma(x)^2}{2}\frac{\partial^2}{\partial u^2} +\gamma(x)\frac{\partial}{\partial u} \left ( u \cdot \right ).\nonumber
\end{equation}
  The relevant (integrable) solution is a mean-zero Gaussian given in Eq. (\ref{eq:invdensity}), which satisfies $g^*(u) = g^*(-u)$, thus 
\begin{equation}
	\int_{-\infty}^\infty g^*L_2g_0\;du = \frac{\partial g_0}{\partial x}\int_{-\infty}^\infty ug^*(u)\;du = 0.\nonumber
\end{equation}
Next the solvability condition for Eq. (\ref{eq:unity}) is 
\begin{equation}
	\label{eq:Lastsolv}
	\int_{-\infty}^\infty \left\{-L_2L_1^{-1}L_2g_0 + \frac{\partial g_0}{\partial t}\right \}g^*\;du = 0.
\end{equation}
First we set
\begin{equation}
    V = L_1^{-1}L_2g_0,\nonumber
\end{equation}
which by the previous solvability condition is well defined.
Thus $L_1V = L_2g_0$, or
\begin{equation}
\label{eq:Vpde}
\frac{\sigma(x)^2}{2}\frac{\partial^2V}{\partial u^2} - \gamma(x) u\frac{\partial V}{\partial u} = u\frac{\partial g_0}{\partial x}.
\end{equation}
Notice that the function
\begin{equation}
    V = -\frac{u}{\gamma(x)}\frac{\partial g_0}{\partial x}\nonumber
\end{equation}
is a particular solution of Eq. (\ref{eq:Vpde}). From Eq. (\ref{eq:Lastsolv}) we must have 
\begin{equation}
	\label{eq:LastsolvInt}
    \int_{-\infty}^\infty g^* \left (\frac{u^2}{-\gamma(x)}\frac{\partial^2 g_0}{\partial x^2} + \frac{u^2}{\gamma(x)^2}\frac{d\gamma(x)}{d x}\frac{\partial g_0}{\partial x} - \frac{F(x)}{\gamma(x)}\frac{\partial g_0}{\partial x} \right )\;du = -\frac{\partial g_0}{\partial t}, 
\end{equation}
for any $g^*$ satisfying $L_1^*g^* =0$, in particular for the invariant density of the OU process $\tilde{u}_t$.
After Gaussian integration over $u$  Eq. (\ref{eq:LastsolvInt}) becomes
\begin{equation}
	\label{eq:genericfricBK}
 \frac{\sigma(x)^2}{2\gamma(x)^2} \frac{\partial^2 g_0}{\partial x^2} + \left (\frac{F(x)}{\gamma(x)} -\frac{\sigma(x)^2}{2\gamma(x)^3}\frac{d\gamma(x)}{dx} \right )\frac{\partial g_0}{\partial x} = \frac{\partial g_0}{\partial t}. 
\end{equation}
This gives the SK approximation to the backward Kolmogorov equation.  The corresponding (It\^o) SDE is 
\begin{equation}
\label{eq:SDEgenericSK}
dx_t = \left (\frac{F(x_t)}{\gamma(x_t)} -\frac{\sigma(x_t)^2}{2\gamma(x_t)^3}\frac{d\gamma(x_t)}{dx} \right )\,dt + \frac{\sigma(x_t)}{\gamma(x_t)}\,d{W}_t.
\end{equation}
{Since we derived this equation from the convergence of the infinitesimal operators, rather than directly studying the limit of the SDE (\ref{eq:SDEgeneral}), the convergence is in law.}

\section{An equation for $\alpha(x)$}\label{sec:alpha}

We derive an equation for $\alpha(x)$ depending on the friction $\gamma(x)$ and diffusion $\sigma(x)$ comparing the backward Kolmogorov equation of the SDE [Eq. (\ref{eq:SKsdeItox})] and Eq. (\ref{eq:SDEgenericSK}) and solving for $\alpha(x)$:
\begin{equation}\label{eq:alpha}
\alpha(x) = \frac{\gamma'(x)\sigma(x)}{2( \gamma'(x)\sigma(x)-\gamma(x)\sigma'(x))},
\end{equation}
where $\gamma'(x) = \frac{d\gamma(x)}{dx}$ and $\sigma'(x) = \frac{d\sigma(x)}{dx}$. This equation shows that in general, $\alpha$ varies with position and can even take values outside the interval $[0,1]$. Interestingly, $\alpha$ never takes the value $\frac{1}{2}$, i.e. we never obtain a Stratonovich correction.

\subsection{$\gamma(x) \equiv \gamma_0$: Constant friction}\label{ssec:constant}

\begin{figure}
\centering
\includegraphics[width=12cm]{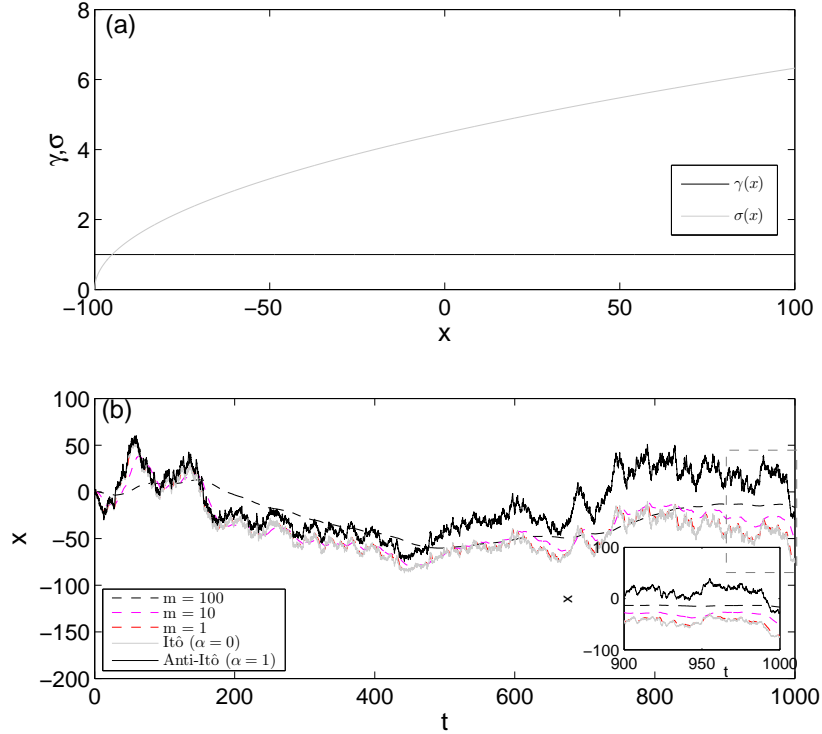}
\caption{(a) $\sigma(x)$ (grey line) and $\gamma(x) = \sigma(x)^0 = \mathrm{constant}$ (dark line). (b) The solutions of the Newton equations [Eq. (\ref{eq:LEgeneral})] for $m \rightarrow 0$ (dashed lines) converge to the solution of the SK approximation [Eq. (\ref{eq:SDE})] for $\alpha = 0$, i.e. It\^o integral, (grey solid line); also the solution for $\alpha = 1$, i.e. anti-It\^o integral, (black solid line) is given for comparison. All solutions are obtained for the same Wiener process. The inset is a blow-up of the final part of the trajectories (dashed square).}
\label{fig:lambda0}
\end{figure}

The case in which $\gamma(x) \equiv \gamma_0$, while $\sigma(x)$ is allowed to vary [Fig. \ref{fig:lambda0}(a)], has been often object of mathematical studies. For example, Freidlin \cite{freidlin2004} and later Pavliotis and Stuart \cite{pavliotis2005} proved that the limiting equation has a stochastic term with $\alpha=0$; this result is rederived here.
Physically, an example of this system is in the framework of the Maxey-Riley model of inertial particles in a Gaussian field \cite{sigurgeirsson2002} with correlation time assumed to be very short. In Fig. \ref{fig:lambda0}(b), we show how the numerical solutions for $m�\rightarrow 0$ converge towards the It\^o ($\alpha = 0$) solution of Eq. (\ref{eq:SDE}).

\subsection{$\gamma(x) \propto \sigma(x)^2$: Brownian motion}\label{ssec:brownian}

The particular case when Einstein fluctuation-dissipation relation is satisfied in its standard form [Eq. (\ref{eq:einstein})] is particularly important because it describes the diffusion of Brownian particles. If $D(x)$ denotes the hydrodynamic diffusion coefficient and $k_B T$ the thermal energy, then 
\begin{equation}
\label{eq:FDgamma}
\gamma(x) = {k_BT \over D(x)}
\end{equation}
and
\begin{equation}
\label{eq:FDsigma}
\sigma(x) = \frac{k_BT\sqrt{2}}{\sqrt{D(x)}}.
\end{equation}
This case was studied experimentally in \cite{volpe2010,brettschneider2011}, showing that the correct value of $\alpha$ for $m \to0$ is $\alpha = 1$. It was subsequently proven by a mathematical argument \cite{wehr2011} that in this case the processes $x_t^m$ converge to the solution of the limiting equation with $\alpha = 1$ in the $L^2$ sense; this result is rederived here, but only in a weaker sense.

\subsection{$\gamma(x) \propto \sigma(x)^\lambda$: Constant $\alpha$}\label{ssec:specialcases}

\begin{figure}
\centering
\includegraphics[width=5in]{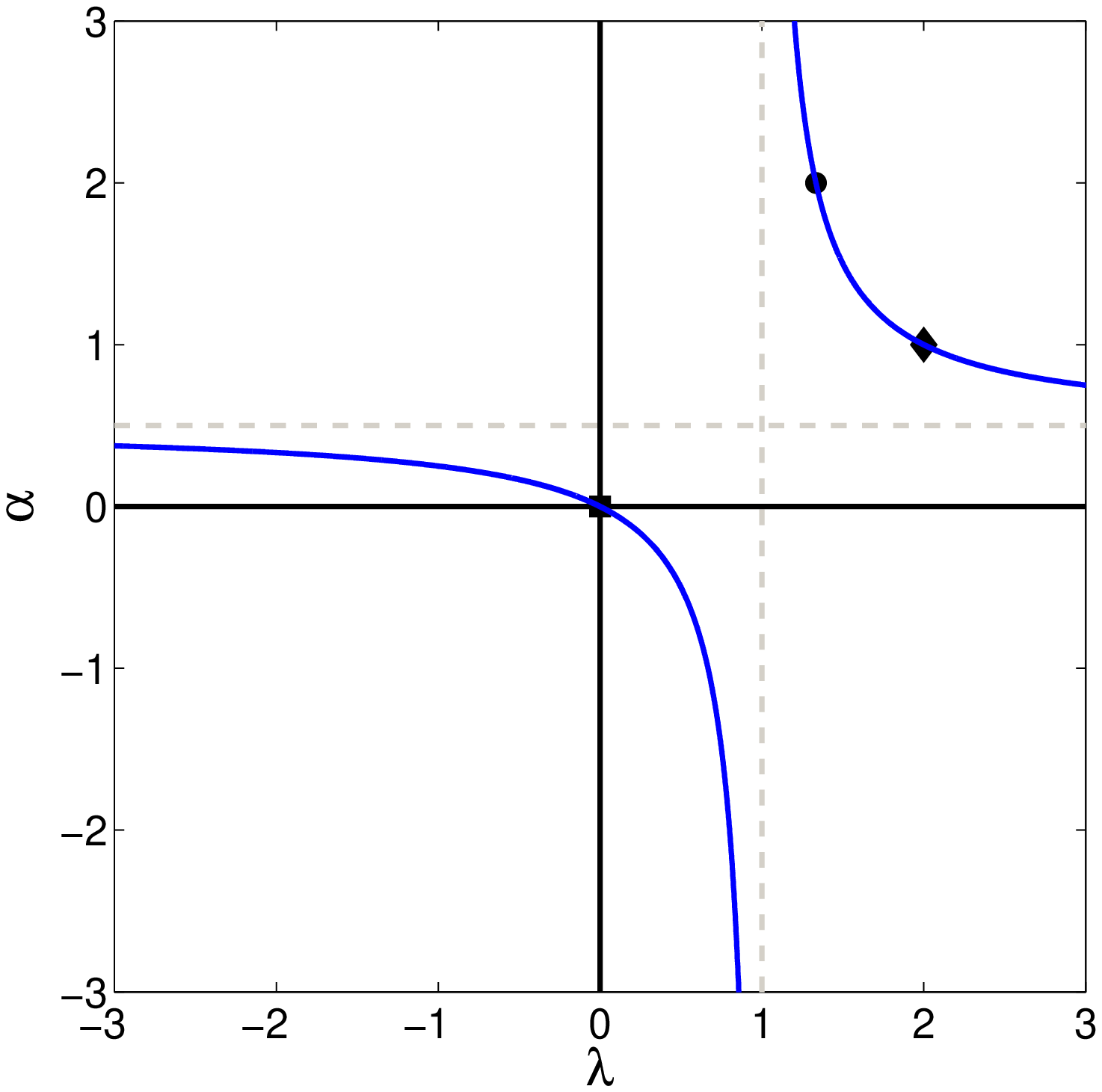}
\caption{$\alpha$ as a function of $\lambda$ for the case when $\gamma(x) \propto \sigma(x)^\lambda$ [$\mathsection$ \ref{ssec:specialcases}]. For $\lambda \rightarrow 1$,$\alpha$ diverges asymptotically (dashed line) leading to the singular case discussed in $\mathsection$ \ref{ssec:proportional}. The It\^o integral ($\alpha = 0$) is obtained for $\lambda = 0$ (square) and the anti-It\^o ($\alpha = 1$) for $\lambda = 2$ (diamond); the Strasonovich integral ($\alpha = 0.5$) is only obtained asymptotically (dotted line) for $\lambda \rightarrow \infty$.}
\label{fig:lambda-alpha}
\end{figure}

\begin{figure}
\centering
\includegraphics[width=5in]{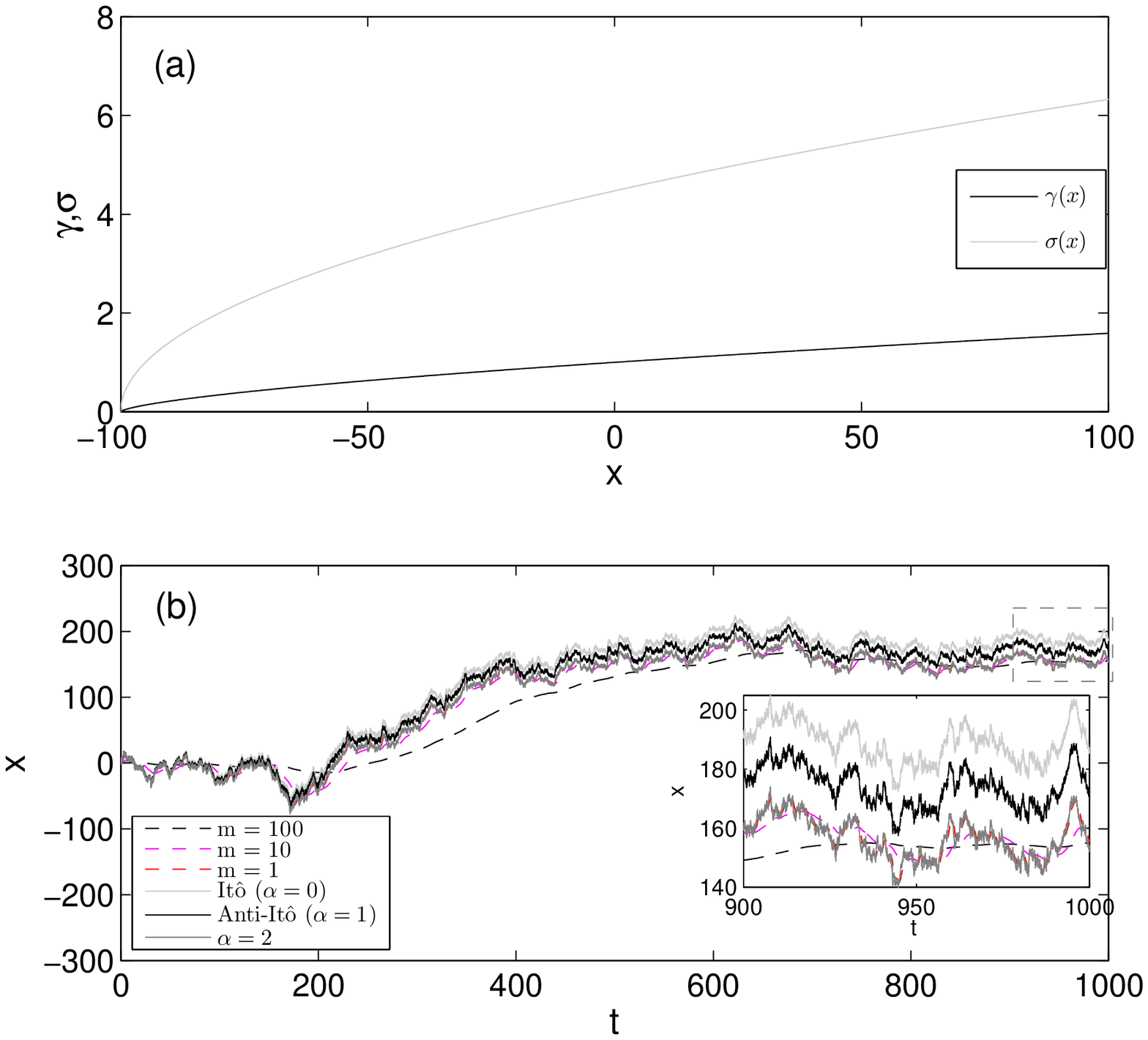}
\caption{(a) $\sigma(x)$ (grey line) and $\gamma(x) = \sigma(x)^{4/3}$ (dark line). (b) The solutions of the Newton equations [Eq. (\ref{eq:LEgeneral})] for $m \rightarrow 0$ (dashed lines) converge to the solution of the SK approximation [Eq. (\ref{eq:SDE})] for $\alpha = 2$, (dark grey solid line); also the solution for $\alpha = 0$, i.e. It\^o integral (grey solid line), and $\alpha = 1$, i.e. anti-It\^o integral, (black solid line) is given for comparison. All solutions are obtained for the same Wiener process. The inset is a blow-up of the final part of the trajectories (dashed square)}
\label{fig:prop_alpha_2}
\end{figure}

All the cases for which $\alpha(x) \equiv \alpha$, can be obtained equating the right-hand side of Eq. (\ref{eq:alpha}) to a constant, different from $\frac{1}{2}$. After a simple calculation, we obtain $\gamma(x) = c\sigma(x)^{\lambda}$, where $c$ is a constant, i.e. Eq. (\ref{eq:generalized}). It follows that
\begin{equation}\label{eq:alpha_lambda}
	\alpha = \frac{\lambda}{2(\lambda - 1)}.
\end{equation}
The value of $\alpha$ depends on the exponent $\lambda$ as shown in Fig. \ref{fig:lambda-alpha}. 
This result includes as particular cases $\gamma(x)\equiv\gamma$, for which $\alpha = 0$ [$\mathsection$ \ref{ssec:constant}], and $\gamma(x) = c\sigma(x)^2$, for which $\alpha = 1$ [$\mathsection$ \ref{ssec:brownian}]. However, we remark that the value $\alpha = {1 \over 2}$ is only obtained asymptotically for $\lambda \rightarrow \infty$.

Interestingly, values of $\alpha$ outside the interval $[0,1]$ can be achieved for certain friction--diffusion relations. For example, the relation $\gamma(x) = \sigma(x)^{4/3}$ gives $\alpha = 2$ by the formula [Eq. (\ref{eq:alpha_lambda})]. Figure (\ref{fig:prop_alpha_2}) gives insight to the zero mass limit. Different constructions of the stochastic integral are given for It\^o (grey solid line), anti-It\^o (black solid line), and for $\alpha = 2$ (dark grey solid line), for the same Wiener process.  

\subsection{$\gamma(x) \propto \sigma(x)$: A singular case}\label{ssec:proportional}

\begin{figure}
\centering
\includegraphics[width=12cm]{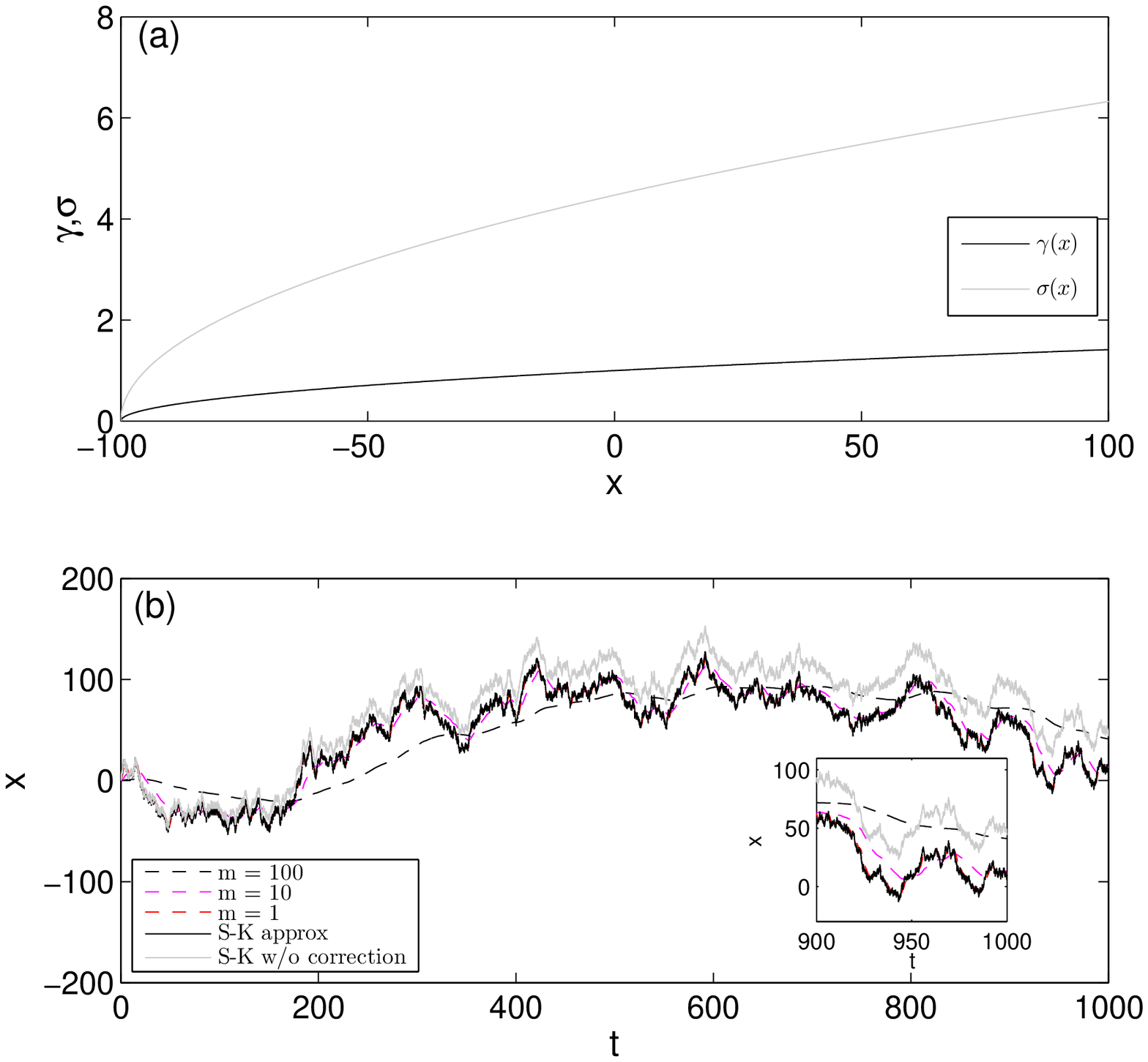}
\caption{(a) $\gamma(x) = c\sigma(x)$. (b) For $m \rightarrow 0$, the solutions of Eq. (\ref{eq:SDEgeneral}) (dashed lines) converge to the limiting Eq. (\ref{eq:propgamma}) (dark solid line). The grey solid line represents the solution of Eq. (\ref{eq:SDE}) disregarding the noise-induced drift. All solutions are obtained for the same Wiener process. The inset is a blow-up of the final part of the trajectories (dashed square).}
\label{fig:prop}
\end{figure}

When $\gamma(x) \propto \sigma(x)$ [Fig. \ref{fig:prop}(a)], the stochastic term in Eq. (\ref{eq:SDE}) gets multiplied by a constant factor, i.e. $\displaystyle \frac{\sigma(x_t)}{\gamma(x_t)} = \mathrm{constant}$, and thus there is no ambiguity in its solution. However, the zero-mass limit of Eq. (\ref{eq:LEgeneral}) does not converge to this solution. 
This can be seen by setting $\gamma(x) = c\sigma(x)$ and using Eq. (\ref{eq:genericfricBK}) directly; the limiting equation is
\begin{equation}
\frac{\partial g_0}{\partial t} = \frac{1}{2c^2}\frac{\partial^2 g_0}{\partial x^2} + \left ( \frac{F(x)}{c\sigma(x)} - \frac{\sigma'(x)}{2c^2\sigma(x)}\right )\frac{\partial g_0}{\partial x}.
\end{equation}
This gives the Smoluchowski-Kramers limiting SDE as
\begin{equation}
	\label{eq:propgamma}
	dx_t = \left (\frac{F(x_t)}{c\sigma(x_t)} - \frac{\sigma'(x_t)}{2c^2\sigma(x_t)}\right )\;dt + \frac{1}{c}\,d\tilde{W}_t.
\end{equation}
Here we see that there is a correction to the drift term. In Fig. \ref{fig:prop}(b), we show how the numerical solutions for $m�\rightarrow 0$ (dashed lines) converge towards the solution of the SK approximation (black solid line), while the solution Eq. (\ref{eq:SDE}) without the correction to the drift clearly diverges (grey solid line).


\section{Conclusion and future research}

We have performed a systematic study of the Smoluchowski-Kramers limit for a class of SDEs with arbitrary friction and diffusion. We have identified the It\^o form of the limiting Langevin equation and discussed its equivalent interpretation in terms of other definitions of stochastic integrals.{We introduced the notation $\circ^{\alpha(x)}\;dW_t$, that is interpreted as an It\^o integral with an additional well-defined drift term. However, we have not given a rigorous mathematical construction of this integral, which is nevertheless of interest. In a future work, we will study the analogous problem for some physically relevant systems driven by colored noise.}

\begin{acknowledgements}
We are thankful to C. Bechinger, L. Helden and T. Brettschneider for the privilege of their collaboration on a closely related project and the physical insight we gained from discussions with them. The multiscale analysis of the Kolmogorov equation was suggested to us by S.R.S. Varadhan.  We also thank A. Danev and A. McDaniel for interesting and useful discussions on the subject of this work and D. Herzog for critical reading of the manuscript. J.W. thanks C. Bechinger and Universit\"at Stuttgart for hospitality.
\end{acknowledgements}

\bibliographystyle{spphys}       
\bibliography{sde_bib2}   

\end{document}